\documentclass[
12pt,
preprint,preprintnumbers,nofootinbib,
groupedaddress,superscriptaddress,amsmath,amssymb]{revtex4}
\usepackage{tabularx}
\usepackage{graphicx}
\usepackage{dcolumn}
\usepackage{bm}
\usepackage{amssymb}
\usepackage{amsmath}
\usepackage{epsfig}    
\usepackage{color}
\usepackage{slashed}
\usepackage{hhline}

\def\be{\begin{equation}}
\def\ee{\end{equation}}
\newcommand{\bea}{\begin{eqnarray}}
\newcommand{\eea}{\end{eqnarray}}
\newcommand{\nn}{\nonumber}

\begin{document}

{\begin{flushright}{APCTP Pre2018 - 019}\end{flushright}}

\title{Resolving $B$-meson  anomalies by flavor-dependent gauged symmetries $\displaystyle\prod_{i=1}^3U(1)_{B_i-L_i}$ }

\author{Chao-Qiang Geng}
\email{geng@phys.nthu.edu.tw}
\affiliation{Chongqing University of Posts \& Telecommunications, Chongqing 400065}
\affiliation{Department of Physics, National Tsing Hua University, Hsinchu  300}
\affiliation{Physics Division, National Center for Theoretical Sciences, Hsinchu 300}

\author{Hiroshi Okada}
\email{hiroshi.okada@apctp.org}
\affiliation{Asia Pacific Center for Theoretical Physics, Pohang, Gyeongbuk 790-784, Republic of Korea}

\date{\today}

\begin{abstract}
We propose a model with flavor dependent gauged symmetries of $\displaystyle\prod_{i=1}^3 U(1)_{B_i-L_i}$ with $i$  the family indices.
After formulating the  renormalizable Yukawa Lagrangian, Higgs potential and kinetic term,
 we study  the lepton sector based on a successful two-zero texture without introducing extra scalar  bosons to avoid the dangerous Goldstone bosons.
In particular, we discuss the muon related phenomenologies via additional neutral gauge bosons.
In our numerical analysis, we explore the allowed parameter space, in which 
the anomaly of $B\to K^*\bar\mu\mu$ can be explained. 
\end{abstract}
 \maketitle
\newpage

\section{Introduction}

Recently, there have been some anomalies in semi-leptonic $B$-meson decays~\cite{Aaij:2014ora,Aaij:2017vbb,Aaij:2015esa,Khachatryan:2015isa,Lees:2015ymt,Wei:2009zv,Aaltonen:2011ja,Aaij:2015oid,Wehle:2016yoi,Sirunyan:2017dhj,Lees:2013uzd,Aaij:2015yra}, 
which would be hints of new physics (NP) beyond the standard model (SM)~\cite{Hurth:2016fbr,DAmico:2017mtc,Altmannshofer:2017yso,Hiller:2017bzc,Geng:2017svp,Ciuchini:2017mik,Celis:2017doq,Hurth:2017hxg,Capdevila:2017bsm}. 
For example, the  ratio of
$R_{K^{*}}\equiv {\cal B}(B^0\to K^{*0}\mu^+\mu^-)/{\cal B}(B^0\to K^{*0}e^+e^-)$ 
has been measured to be~\cite{Aaij:2017vbb}
\bea\label{R_K*}
R_{K^*}^{\rm expt} &=&
\left\{
\begin{array}{cc}
0.66^{+0.11}_{-0.07}~{\rm (stat)} \pm 0.03~{\rm (syst)} ~,~~ & 0.045 \le q^2 \le 1.1 ~{\rm GeV}^2 ~, \\
0.69^{+0.11}_{-0.07}~{\rm (stat)} \pm 0.05~{\rm (syst)} ~,~~ & 1.1 \le q^2 \le 6.0 ~{\rm GeV}^2 ~,
\end{array}
\right.
\eea
where $q^2$ is the invariant mass for the final lepton pair. 
In the SM, 
$R_{K^{*}}$ 
is expected to be close to 1~\cite{Hiller:2003js,Bordone:2016gaq}.
It has been shown that these anomalous results can be explained by NP based on 
the model-independent studies, such as those in Refs.~\cite{Hurth:2016fbr,DAmico:2017mtc,Altmannshofer:2017yso,Hiller:2017bzc,Geng:2017svp,Ciuchini:2017mik,Celis:2017doq,Hurth:2017hxg,Capdevila:2017bsm}. 
On the other hand, it has been anticipated that these anomalies arise from flavor-dependent effects~\cite{Ko:2017quv,Ko:2017yrd,Bordone:2017anc,Bonilla:2017lsq,Tang:2017gkz,Mu:2018weh,Chen:2017usq,Bian:2017rpg,Allanach:2018lvl,Kamada:2018kmi} due to the violation of the lepton flavor universality.

In our previous paper~\cite{Mu:2018weh}, we  have shown several phenomenological insights based on flavor-dependent gauged symmetries
of $U(1)_{B-L_1}\times U(1)_{B-L_{2}-L_{3}}$,
in which we have explored the Yukawa sector by introducing additional Higgs bosons to evade the dangerous goldstone bosons (GBs)
in order to understand the anomaly of $B\to K^*\mu\bar\mu$~\cite{Aaij:2017vbb,Hurth:2016fbr}.
In this study, we further extend the flavor-dependent gauge symmetry into $U(1)_{B_1-L_1}\times U(1)_{B_2-L_2}\times U(1)_{B_3-L_3}$.
As a result, we can successfully resolve the anomaly of $B\to K^*\mu\bar\mu$ without  adding any new fields besides right-handed neutrinos,
while the dangerous GB can naturally be evaded.
Furthermore, a see-saw type of neutrino masses can be realized  in the lepton sector with a specific two-zero texture~\cite{Fritzsch:2011qv}.

This paper is organized as follows.
In Sec.~II, we first describe our field contents along with their charge assignments and  write down the renormalizable Lagrangian
 with the Yukawa integration as well as  Higgs  and neutral vector gauge boson sectors.
We then formulate the mass matrix for the quark and lepton sectors,
in which we concentrate on the predictions in the lepton sector. 
After that, we discuss muon related phenomenologies in the additional neutral gauge bosons,
in which we write down the relevant Lagrangian, the formulas  for $B\to K^*\mu\bar\mu$ and the meson mixings of $M-\bar M$ ($M=K^0,B_d,B_s$), 
and the bound from the LHC data.
In Sec.~III,  we perform a numeral analysis and show the allowed region to satisfy the anomaly without conflict of the constraints. 
Finally, we  conclude in Sec.~IV with some discussions.

\section{ Model setup and phenomenology}
We extend the flavor-blind gauge symmetry in the SM by
 imposing three additional flavor-dependent $U(1)_{B_i-L_i}$ (i=1,2,3)
 gauge groups, with including three right-handed neutral fermions $N_{R_{1,2,3}}$, where  the subscripts represent the family indices.
The field contents of fermions (scalar bosons) under the symmetries of $U(1)_{B_i-L_i}$ (i=1,2,3)
($SU(2)_L\times U(1)_Y\times \displaystyle\prod_{i=1}^3U(1)_{B_i-L_i})$
are given in Table~\ref{tab:1} (\ref{tab:2}).
The anomaly cancellations among $U(1)_Y$ and $U(1)_{B_i-L_i}$ (i=1,2,3)
are straightforwardly derived similar to those in ref.~\cite{Mu:2018weh}.
In Table~\ref{tab:2},
$H_0$ is expected to be the SM Higgs,
while $H_{i}$ (i=1,2,3) are the new isospin doublet scalar bosons, which play a role in providing the mixings of the 1-2 ,2-3 and 1-3 components
for the down quark sector. Under these symmetries, the renormalizable Yukawa Lagrangian is given by 
\begin{align}
-{\cal L}=&
 \sum_{i=1,2,3}\left(y_{u_i} \bar Q_{L_i}\tilde H_0 u_{R_i}  +y_{d_f} \bar Q_{L_i} H_0 d_{R_i} 
 +y_{\nu_i} \bar Q_{L_i}\tilde H_0 N_{R_i}  + y_{\ell_i} \bar L_{L_i} H_0 e_{R_i}\right) 
 \nonumber \\
&+y_{d_{12}}\bar Q_{L_1} H_1 d_{R_2}  +y_{d_{13}} \bar Q_{L_1} H_2 d_{R_3}  +y_{d_{23}} \bar Q_{L_2} H_3 d_{R_3}
\nn\\
&+ \frac12 y_{N_{1}}\varphi_{0}\bar N^C_{R_1} N_{R_1}+ \frac12 y_{N_{2}}\varphi_{1}\bar N^C_{R_2} N_{R_2}\nn\\
& + \frac12 y_{N_{12}}\varphi_{2} (\bar N^C_{R_1} N_{R_2} +\bar N^C_{R_2} N_{R_1}) 
  + \frac12 y_{N_{23}}\varphi_{3} (\bar N^C_{R_2} N_{R_3} +\bar N^C_{R_3} N_{R_2}) 
+{\rm h.c.},
\label{eq:majo}
\end{align}
where $\tilde H \equiv (i \sigma_2) H^*$ with $\sigma_2$ being the second Pauli matrix.
From the above Lagrangian, 
one finds that 
 the Cabibbo-Kobayashi-Maskawa (CKM)~\cite{ckm}
quark  and Pontecorvo-Maki-Nakagawa-Sakata (PMNS)~\cite{PMNS} lepton
mixing matrices  arise only from the down-quark and  neutrino sectors  due to their diagonal up-quark and charged-lepton terms, 
assured by our additional gauge symmetries.

\begin{center} \begin{table}[t]\begin{tiny}
\caption{Field contents of fermions 
and their charge assignments under $U(1)_{B_i-L_i}$ (i=1,2,3), where the subscripts
 $i$ correspond to the family indices.}
 \vspace{0.5cm}
 \begin{tabular}{|c||c|c|c|c|c|c|c|c|c|c|c|c|c|c|c|c|c|c|c|c}
\hline
Fermions& ~$Q_{L_1}$~ & ~$Q_{L_2}$~  & ~$Q_{L_3}$~  & ~$u_{R_1}$~& ~$u_{R_2}$~& ~$u_{R_3}$~ & ~$d_{R_1}$~ & ~$d_{R_2}$~& ~$d_{R_3}$~ &~$L_{L_1}$~&~$L_{L_2}$~&~$L_{L_3}$~ 
& ~$e_{R_1}$~ & ~$e_{R_2}$~& ~$e_{R_3}$~ & ~$N_{R_1}$~ & ~$N_{R_2}$~ & ~$N_{R_3}$~ 
\\
\hline 
\hline
 $U(1)_{B_1-L_1}$ & $\frac13$ & $0$& $0$  & $\frac13$ & $0$ & $0$ & $\frac13$ &  $0$ & $0$ & $-1$& $0$ & $0$ & $-1$  & $0$ & $0$ & $-1$  & $0$  & $0$ \\\hline
  $U(1)_{B_2-L_2}$  & $0$ & $\frac13$  & $0$  & $0$ & $\frac13$& $0$  & $0$ & $\frac13$ & $0$ & $0$  & $-1$   & $0$  & $0$  & $-1$ & $0$  & $0$  & $-1$  & $0$   \\\hline
  $U(1)_{B_3-L_3}$  & $0$ & $0$  & $\frac13$ & $0$   & $0$ & $\frac13$& $0$ & $0$ & $\frac13$  & $0$ & $0$ & $-1$  & $0$   & $0$  & $-1$  & $0$  & $0$  & $-1$   \\\hline
\end{tabular}\label{tab:1} \end{tiny}\end{table}\end{center}

\begin{table}[t]
\caption{Field contents of scalar bosons 
and their charge assignments under  $SU(2)_L\times U(1)_Y\times U(1)_{B_1-L_1}\times U(1)_{B_2-L_2}\times U(1)_{B_3-L_3}$, where all of the scalar fields are singlet under $SU(3)_C$.}
\vspace{0.5cm}
\centering {\fontsize{10}{12}
\begin{tabular}{|c||c|c|c|c||c|c|c|c|}\hline
  Bosons  &~ $H_0$~ &~ $H_1$~ &~ $H_2$~ &~ $H_3$~ &~$\varphi_{0}$~&~$\varphi_{1}$~&~$\varphi_{2}$~&~$\varphi_{3}$~  \\\hline
\hline 
$SU(2)_L$ & $\bm{2}$& $\bm{2}$ & $\bm{2}$& $\bm{2}$ & $\bm{1}$& $\bm{1}$& $\bm{1}$& $\bm{1}$ \\\hline 
$U(1)_Y$ & $\frac12$ & $\frac12$ & $\frac12$ & $\frac12$ & $\bm{0}$  & $\bm{0}$& $\bm{0}$  & $\bm{0}$ \\\hline
 $U(1)_{B_1-L_1}$ & $0$  & $\frac13$& $\frac13$  & $0$     & $2$   & $0$   & $1$    & $0$ \\\hline
 $U(1)_{B_2-L_2}$ & $0$  & $\frac13$&   $0$ & $-\frac13$     & $0$   & $2$   & $1$    & $1$ \\\hline
  $U(1)_{B_3-L_3}$ & $0$  &   $0$ & $-\frac13$ & $-\frac13$     & $0$   & $0$   & $0$    & $1$ \\\hline
  \end{tabular}%
} 
\label{tab:2}\end{table}

\noindent \underline{\it Scalar sector}:
The renormalizable Higgs potential is given by
\begin{align}
V&= \sum_{i=0,1,2,3}\left({\mu_{H_i}^2} |H_i|^2 + {\mu_{\varphi_i}^2} |\varphi_i|^2 \right)
 \nn\\
&+
 \sum_{i=0,1,2,3} \left({\lambda_{H_i}} |H_i|^4 + {\lambda_{\varphi_i}} |\varphi_i|^4   \right)
 + \sum_{i> j =0,1,2,3} \left({\lambda_{H_{ij}}} |H_i|^2  |H_j|^2 + {\lambda'_{H_{ij}}} |H_i^\dag H_j|^2 +{\lambda_{\varphi_{ij}}} |\varphi_i|^2  |\varphi_j|^2 \right)\nn\\
 &+\lambda_0(H^\dag_1 H_2)(H^\dag_3 H_0)
+\lambda'_0(H^\dag_1 H_3)(H^\dag_0 H_2)
+\lambda''_0(H^\dag_2 H_3)(H^\dag_0 H_1) 
+\lambda'''_0(\varphi_{0} \varphi_{1} )\varphi_{2}^{*2} + {\rm h.c.}
,
\label{eq:lag-lep}
\end{align}
where we have neglected the mixing terms between $H_i$ and $\varphi_j$ for simplicity, and $\lambda_0^\alpha$ are non-trivial terms that can forbid 
the dangerous GB.
The scalar fields are parameterized as 
\begin{align}
&H_i =\left[\begin{array}{c}
w^+_i\\
\frac{v_i + h_i +i z_i}{\sqrt2}
\end{array}\right],\quad  
\varphi_i=
\frac{v'_{i} +\varphi_{R_i} + iz'_i}{\sqrt2},\ (i=0,1,2,3),
\label{component}
\end{align}
where one of the mass eigenstates  of $z_{0,1,2,3}$ ($w^\pm_{0,1,2,3}$)
is absorbed by the SM  vector gauged boson $Z$ ($W^\pm$),
while  three of the mass eigenstates of $z'_{0,1,2,3}$  by the additional vector gauged bosons
 $Z^{\prime,\prime\prime,\prime\prime\prime}$, respectively.
 Here, $Z\equiv (g_1^2+g_2^2)v/4$ with $v\equiv \sqrt{v_0^2+  v_1^2+  v_2^2+  v_3^2}\approx 246$ GeV and $Z^{\prime,\prime\prime,\prime\prime\prime}$ arise from $U(1)_{B_i-L_i}$ as we will see later.
Here, we just write down the massive eigenvalue of the CP-odd boson in $\varphi_j$:
\begin{align}
& m_{z'}^2 
=\lambda'''_0
\frac{(v'^2_3+v'^2_1)v'^2_0 +  v'^2_1 v'^2_3}{2 v'_0 v'_1},
\end{align}
where we have defined $O_{z'} M^2_{z'} O_{z'}^T={\rm diag}[0,0,0,m_{z'}^2]$ with $O_{z'}$  the
orthogonal mixing matrix and $M_{z'}$  the four by four symmetric mass matrix among $z'$.  
Similar to the above, we also formulate the other sectors as follows: $O_{z} M^2_{z} O_{z}^T={\rm diag}[0,m^2_{z_1},m^2_{z_2},m^2_{z_3}]$, $O_{h} M^2_{h} O_{h}^T={\rm diag}[m^2_{h_0},m^2_{h_1},m^2_{h_2},m^2_{h_3}]$, $O_{\varphi_R} M^2_{\varphi_R} O_{\varphi_R}^T={\rm diag}[m^2_{\varphi_0},m^2_{\varphi_1},m^2_{\varphi_2},m^2_{\varphi_3}]$, and $O_{w} M^2_{w^\pm} O_{w}^T={\rm diag}[0, m^2_{w_1},m^2_{w_2},m^2_{w_3}]$.
{\it Remarkably, we do not need any additional Higgs bosons to forbid the dangerous GB in spite of many Higges! }

 \subsection{Neutral gauge boson sector}
 
 \noindent
\underline{\it $Z$-$Z'$-$Z''$-$Z'''$ mixing}:
Here, we describe  the neutral gauge bosons among  $Z$-$Z'$-$Z''$-$Z'''$.
But once  $v_i<<v'_i$ (i=0-3) are assumed, $Z$ and $Z^{\prime,\prime\prime,\prime\prime\prime}$  can be decomposed.
Consequently, we can choose $Z$ as the SM gauge boson, while $Z^{\prime,\prime\prime,\prime\prime\prime}$  as the new ones, separately.
Below, we consider the new gauge sector. 
The resulting mass matrix in the basis of $(Z',Z'',Z''')$  is  given by
\begin{align}
m_{Z',Z'',Z'''}^2
&= 
\left[\begin{array}{ccc}
g'^2_1 (4 v'^2_0 + v'^2_2) & g'_1 g'_2 v'^2_2 & 0  \\ 
 g'_1 g'_2 v'^2_2 &  g'^2_2 (4 v'^2_1 + v'^2_2 + v'^2_3)  & g'_2 g'_3 v'^2_3   \\
0  &    g'_2 g'_3 v'^2_3  & g'^2_3 v'^2_3   \\ 
\end{array}\right],
\end{align}  
where  $g'_i$ (i=1,2,3) are the new gauge couplings under $U(1)_{B_i-L_i}$ (i=1,2,3), respectively.
Here, we further impose an assumption $v'_2<< v'_{0,1,3}$ in our convenience later.
In this case, {\it one of the mass eigenstates is uniquely fixed to be $m^2_{Z'_1}\approx 4g'^2_1 v'^2_0$ that can be regarded as an electron specific vector gauge boson, and its mass is expected to be very large through experiments such  as LEP and LHC.} 
Subsequently, the reduced mass matrix is given by
\begin{align}
m_{Z'',Z'''}^2
&\sim g'^2_3 v'^3_3
\left[\begin{array}{cc}
(1+4r)\epsilon^2 & \epsilon  \\
\epsilon  & 1   \\ 
\end{array}\right], \quad \left(r\equiv \frac{v'^2_1}{v'^2_3},\quad \epsilon\equiv\frac{g'_2}{g'_3}\right)
\end{align}  
{
which is diagonalized by the two-by-two mixing matrix $V_G$ as $V_G m_{Z'',Z'''}^2 V_G^T
\equiv {\rm Diag}(m^2_{Z'_2},m^2_{Z'_{3}})$ with 
 \begin{align}
m^2_{Z'_2}&=\frac{g'^2_3 v'^3_3}2
\left[1+(1+4r)\epsilon^2 - \sqrt{1+2(1-4 r)\epsilon^2+(1+4r)^2\epsilon^4 } \right],\nn\\
m^2_{Z'_3} &=\frac{g'^2_3 v'^3_3}2
\left[1+(1+4r)\epsilon^2 + \sqrt{1+2(1-4 r)\epsilon^2+(1+4r)^2\epsilon^4 } \right],\\
V_G&= 
\left[\begin{array}{cc}
c_\theta&    -s_\theta \\
s_\theta & c_\theta  \\ 
\end{array}\right],\quad 
t_{2\theta} = \frac{2\epsilon}{(1+4r)\epsilon^2-1}
\label{eq.st}.
\end{align}

\noindent \underline{\it Fermion sector}:
The mass matrices for the quark sector are given by
 \begin{align}
M_u&= 
\left[\begin{array}{ccc}
m_u &  0&  0 \\
0 &  m_c &  0 \\
0 &  0&  m_t \\
\end{array}\right],
\quad 
M_d=
\left[\begin{array}{ccc}
m_d &  m_{ds} &  m_{db} \\
0 &  m_s &  m_{sb} \\
0 &  0&  m_b \\
\end{array}\right]
\label{eq:q-mass},
\end{align}  
where $m_{u,c,t}\equiv y_{u_{1,2,3}} v_0/\sqrt{2}$,  $m_{d,s,b}\equiv y_{d_{1,2,3}} v_0/\sqrt{2}$, $m_{ds}\equiv y_{12} v_1/\sqrt2$, $m_{db}\equiv y_{13} v_2/\sqrt2$, and $m_{sb}\equiv y_{23} v_3/\sqrt2$.
It suggests that the observed mixing matrix comes from the down sector;  $V_{CKM} = V_{dL}$, where
the mass matrix for the down sector is diagonalized by bi-unitary mixing matrices as $D_{d}=V_{dL} M_d V_{dR}^\dag$.
Therefore, $|D_d|^2=V_{CKM} M_d M_d^\dag V_{CKM}^\dag$.\\

\begin{figure}[t]
\begin{center}
\includegraphics[width=70mm]{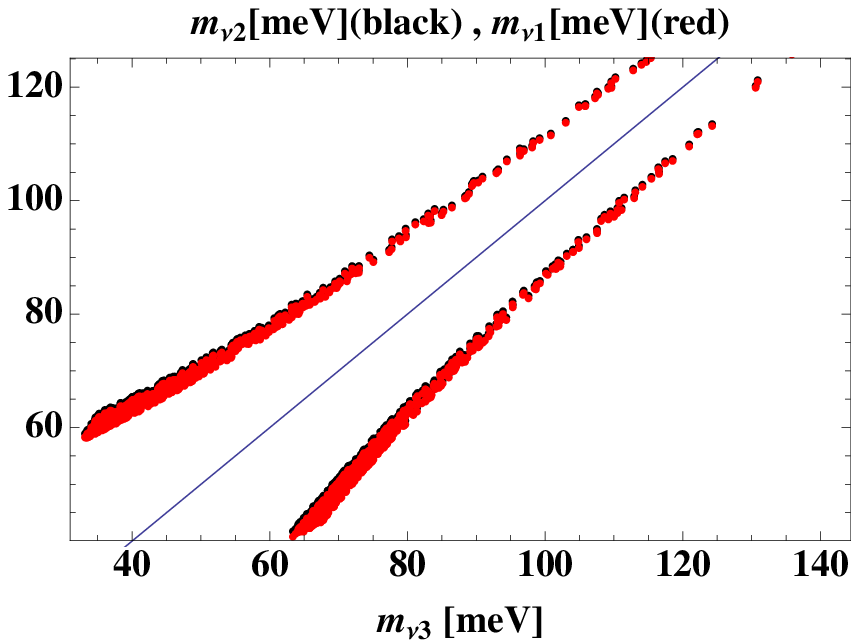} 
\includegraphics[width=70mm]{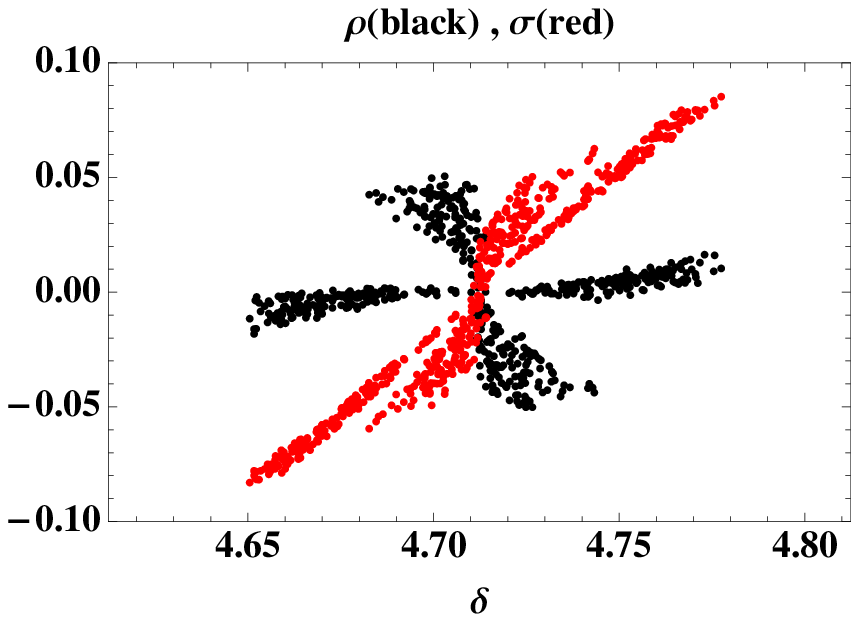} 
\caption{Allowed regions in the planes of  $m_{\nu_3}$-$m_{\nu_{1,2}}$ meV (left) and $\delta$-$(\rho,\sigma)$  (right),
where $\delta$ and $(\rho,\sigma)$ are the Dirac and Majorana CP phases, respectively.} 
  \label{fig:neut}
\end{center}\end{figure}
The mass matrices for the lepton sector are given by
 \begin{align}
M_\ell&= 
\left[\begin{array}{ccc}
m_e &  0&  0 \\
0 &  m_\mu &  0 \\
0 &  0&  m_\tau \\
\end{array}\right],
\quad 
M_D=
\left[\begin{array}{ccc}
m_{D_1} &  0 &  0 \\
0 &  m_{D_2} &  0 \\
0 &  0&  m_{D_3} \\
\end{array}\right],
\quad 
M_N=
\left[\begin{array}{ccc}
m_{N_{11}} &  m_{N_{12}} &  0 \\
m_{N_{12}} &   m_{N_{22}} &   m_{N_{23}} \\
0  &  m_{N_{23}}  & 0 \\
\end{array}\right]
\label{eq:lep-mass},
\end{align}  
where $m_{e,\mu,\tau}\equiv y_{\ell_{1,2,3}} v_0/\sqrt{2}$,  $m_{D_{1,2,3}}\equiv y_{\nu_{1,2,3}} v_0/\sqrt{2}$,
$m_{N_{11}} \equiv y_{N_1} v'_0/\sqrt2$, 
$m_{N_{22}} \equiv y_{N_2} v'_1/\sqrt2$, $m_{N_{12}} \equiv y_{N_{12}} v'_2/\sqrt2$, 
$m_{N_{23}} \equiv y_{N_{23}} v'_3/\sqrt2$.
After applying the seesaw mechanism, the active neutrino mass matrix  $m_\nu$ is given by 
\begin{align}
m_\nu\approx - M_D^T M^{-1}_N M_D =
\left[\begin{array}{ccc}
\times &  0&  \times  \\
0 &  0 &  \times\\
\times & \times & \times \\
\end{array}\right],
\end{align}
where $m_\nu$ is defined by the diagonalized matrix of  $D_{\nu}=U_{MNS}^\dag m_\nu U^*_{MNS}$.
Clearly, the active neutrino mass matrix is a successful two-zero texture that provides some predictions in fig.~\ref{fig:neut},
and the formulas are simply found by directly solving the following two relations~\cite{Fritzsch:2011qv}:
\begin{align}
(m_{\nu})_{12}=(U_{MNS} D_\nu U^T_{MNS})_{12}=0,\quad (m_{\nu})_{22}=(U_{MNS} D_\nu U^T_{MNS})_{22}=0,
\end{align}
where we have used the experimental values at 3$\sigma$ confidential level (C.L.)~\cite{Forero:2014bxa}.
The left plane in fig.~\ref{fig:neut} represents the allowed region of $m_{\nu_{1}}$(red) and $m_{\nu_2}$(black) in terms of $m_{\nu_3}$, where the
unit is meV.
It suggests that this texture allows both of normal  and inverted hierarchies, given by
 65 meV$\lesssim m_{\nu_3}\lesssim$125 meV and   30 meV$\lesssim m_{\nu_3}\lesssim$115 meV, respectively.
The right figure demonstrates the Majorana phases $(\rho,\sigma)$ in terms of the Dirac one $\delta$.
It implies that the Dirac phase is predicted to be $4.65\lesssim\delta\lesssim4.78$, $-0.05\lesssim \rho\lesssim 0.05$ and $-0.09\lesssim \sigma\lesssim 0.09$ at 3$\sigma$ C.L..
Therefore, $\delta\approx 3\pi/2$ is found to be  the best fit value.

\subsection{Muon related phenomenologies}
We now focus on the interactions between fermions and new  gauge bosons $(Z'_2,Z'_3)$.
Since the masses of $Z'_{2,3}$ are not seriously constrained  by the LEP or LHC experiment, 
they do not couple to the electron/positron.
The relevant interacting Lagrangian is given by
\begin{align}
{\cal L}&\sim \frac{Z'^\mu_2}{3} \bar d_{L_i} \gamma_\mu d_{L_j}
\left[ g'_2 c_\theta (V^\dag_{CKM})_{d_i c}(V_{CKM})_{cd_j} - g'_3 s_\theta (V^\dag_{CKM})_{d_i t}(V_{CKM})_{t d_j} \right] \nn\\
&+
\frac{Z'^\mu_3}{3} \bar d_{L_i} \gamma_\mu d_{L_j}
\left[ g'_2 s_\theta (V^\dag_{CKM})_{d_i c}(V_{CKM})_{cd_j} + g'_3 c_\theta (V^\dag_{CKM})_{d_i t}(V_{CKM})_{t d_j} \right] \nn\\
&-
g'_2 \bar \mu\gamma_\mu\mu(c_\theta Z'^\mu_2 +s_\theta Z'^\mu_3)\\
&\equiv
\frac{Z'^\mu_2}{3} \bar d_{L_i} \gamma_\mu d_{L_j} {\cal O}_{2_{d_id_j}}
+
\frac{Z'^\mu_3}{3} \bar d_{L_i} \gamma_\mu d_{L_j} {\cal O}_{3_{d_id_j}}
-
g'_2 c_\theta Z'^\mu_2 \bar \mu\gamma_\mu\mu
-
g'_2 s_\theta Z'^\mu_3 \bar \mu\gamma_\mu\mu
.
\end{align}

\noindent
\underline{\it $B\to K^*\mu\bar\mu$}:
The effective Lagrangian to explain the $B\to K^*\mu\bar\mu$ anomaly is given by
\begin{align}
i {\cal L}\approx i \frac{g'_2}3
\left(\frac{ c_\theta {\cal O}_{2_{sb}}  }{m_{Z'_2}^2} + \frac{s_\theta{\cal O}_{3_{sb}}  }{m_{Z'_3}^2}\right)
 (\bar s\gamma^\mu P_L b)(\bar\mu\gamma_\mu\mu)\,,
\end{align}
which leads to the $\Delta C_9^\mu$ operator to be
\begin{align}
\Delta C_9^\mu\approx \frac{\sqrt2 \pi g'_2}{\alpha_{em} G_F (V_{CKM})_{tb}  (V^*_{CKM})_{ts} } 
\left(\frac{c_\theta {\cal O}_{2_{sb}}  }{m_{Z'_2}^2} + \frac{s_\theta {\cal O}_{3_{sb}}  }{m_{Z'_3}^2}\right).
\end{align}
The most recent global fit of $\Delta C_9^{\mu}$ can be found in Ref.~\cite{Arbey:2019duh}, given by
\begin{equation}
{ -\Delta C_9^{\mu }} = 1.03\pm0.20 .
\label{eq:C9-fit}
\end{equation}
{\it Remarkably, there are no additional constraints,  such as those from $B_{s/d}\to \bar\mu\mu$ 
and ${\rm BR}(B \to K \mu^+ \mu^-)/{\rm BR}(B \to K e^+e^-)$, since the  Lagrangian does not contain the $\Delta C_{10}$ operator and $Z'_2$ does not directly interact with the electron/positron.}

\noindent \underline{\it $M-\overline M$ meson mixings}:  
The extra gauge boson induces the neutral meson mixings of $M-\bar M$ at the tree level,
where $M=(K^0,B_d,B_s)$.
The formulas for the mass splittings are given by~\cite{Gabbiani:1996hi}
\begin{align}
 \Delta m_{K}^{0} \approx & \frac{2}{3}
\left(\frac{ |{\cal O}_{2_{sd}}|^2  }{m_{Z'_2}^2} + \frac{|{\cal O}_{3_{sd}}|^2  }{m_{Z'_3}^2}\right)
m_{K} f^2_{K},\\
 \Delta m_{B_d}^{0} \approx &\frac{2}{3}
\left(\frac{ |{\cal O}_{2_{bd}}|^2  }{m_{Z'_2}^2} + \frac{|{\cal O}_{3_{bd}}|^2  }{m_{Z'_3}^2}\right)
m_{B_d} f^2_{B_d},\\
 \Delta m_{B_s}^{0} \approx &\frac{2}{3}
\left(\frac{ |{\cal O}_{2_{bs}}|^2  }{m_{Z'_2}^2} + \frac{|{\cal O}_{3_{bs}}|^2  }{m_{Z'_3}^2}\right)
m_{B_s} f^2_{B_s},
\label{eq:kk}
\end{align}
which should be less than the experimental values of 
$(3.48\times10^{-4},3.33\times10^{-2},1.17)\times10^{-11}$ GeV~\cite{Olive:2016xmw}, where 
$f_M=(156,191,200)$ MeV and $m_M=(0.498,5.280,5.367)$ GeV, respectively.

\noindent \underline{\it Bound from the LHC}:  
The data from the LHC experiments~\cite{Aaboud:2017buh} 
also restrict the ratio between the extra gauge couplings and their masses.
Here, we estimate them by applying the effective Lagrangian with the resulting relation, give by
\begin{align}
\frac{(30\ {\rm TeV})^2}{12\pi}\lesssim \frac{1}{g'_2}
\left(\frac{c_\theta {\cal O}_{2_{dd}}  }{m_{Z'_2}^2} + \frac{s_\theta {\cal O}_{3_{dd}}}{m_{Z'_3}^2}\right)^{-1} .
\end{align}
This constraint will be taken into consideration in the numerical analysis.

Before showing numerical analysis, we  discuss the relations between $B\to K^*\mu\bar\mu$ and the mixings of $M-\bar M$,
especially $\Delta B_{d,s}$, involving  the bottom quark, which are  strongly constrained  by the experimental data.
Consequently,  one finds that
$\Delta C_9\approx0.1$ at most, which is smaller than the  value in Eq.~(\ref{eq:C9-fit}) by one order of magnitude.
To enhance $\Delta C_9$,
we can introduce one set of vector quarks: $Q'\equiv [U',D']^T$ and $d'$, which are $SU(2)_L$ doublet and  singlet, respectively,
along with one complex boson inert $S$.
 The additional gauged charges assigned as (0,-1/3,0)  and (0,2/3,0) for  $(Q',d')$ and $S$, respectively. 
Then, we write the new part of the Lagrangian as
\begin{align}
-{\cal L}&= f_2 \bar Q_{L_2} Q'_R S + g_2 \bar d_{R_2} d'_L S + M' \bar Q' Q'+ m' \bar d'd' + m_S^2 |S|^2 +{\rm h.c.}\nn\\
&\rightarrow F^\dag_{i2} \bar d_{L_i} d'_R S + G^\dag_{i2} \bar d_{R_i} d'_L S  + M' \bar D' D'+ m' \bar d'd' + m_S^2 |S|^2 +{\rm h.c.},
\end{align}
where $F^\dag_{i2}(G^\dag_{i2})\equiv V^\dag_{CKM_{i2}} f_2(g_2)$.
Here, we have neglected the mass term $\Bar Q' d'$ and additional Higgs potential related to $S$
as well as  and the diagonal up-quark sector for simplicity. 
Subsequently, we find that  the $M-\bar M$ mixings from the box diagrams 
 are given by
\begin{align}
\label{eq24}
 \Delta m_{K}^{1} \approx & \frac{m_K f_K^2}{3(4\pi)^2}
 \left[(F_{22}^* F_{21})^2 F_{box}(m_S,M') +(G_{21}^* G_{22})^2 F_{box}(m_S,m')  \right],\\
\label{eq25}
 \Delta m_{B_d}^{1} \approx & \frac{m_{B_d} f^2_{B_d}}{3(4\pi)^2}
 \left[(F_{23}^* F_{21})^2 F_{box}(m_S,M') +(G_{21}^* G_{23})^2 F_{box}(m_S,m')  \right],\\
 \label{eq26}
 \Delta m_{B_s}^{1} \approx &\frac{m_{B_s} f^2_{B_s}}{3(4\pi)^2}
 \left[(F_{23}^* F_{22})^2 F_{box}(m_S,M') +(G_{22}^* G_{23})^2 F_{box}(m_S,m')  \right],\\
 F_{box}(m_1,m_2)&=\frac{m_1^2-m_2^2+ m_2^2 \ln\left(\frac{m_2^2}{m_1^2}\right)}{(m_1^2-m_2^2)^2}.
\label{eq:kk-box}
\end{align}
When $F_{21,22}$ and $G_{21}$ are taking to be pure imaginary, $G_{23}\approx0$, and the others are real, 
Eqs.~(\ref{eq24})-(\ref{eq26}) can be simplified as
\begin{align}
 \Delta m_{K}^{1} \approx & \frac{m_K f_K^2}{3(4\pi)^2}
 \left[|F_{22}|^2 |F_{21}|^2 F_{box}(m_S,M') -|G_{21}|^2 |G_{22}|^2 F_{box}(m_S,m')  \right],\\
 \Delta m_{B_d}^{1} \approx & -\frac{m_{B_d} f^2_{B_d}}{3(4\pi)^2}
 |F_{23}|^2 |F_{21}|^2 F_{box}(m_S,M') ,\\
 \Delta m_{B_s}^{1} \approx & -\frac{m_{B_s} f^2_{B_s}}{3(4\pi)^2}
 |F_{23}|^2 |F_{22}|^2 F_{box}(m_S,M')\,,
\label{eq:kk-box-approx}
\end{align}
respectively,
leading to 
\begin{align}
 \Delta m_{K}^{total} &\approx \Delta m_{K}^{1}  \approx \frac{m_K f_K^2}{3(4\pi)^2}
 \left[|F_{22}|^2 |F_{21}|^2 F_{box}(m_S,M') -|G_{21}|^2 |G_{22}|^2 F_{box}(m_S,m')  \right],\label{eq:kk-total-k}
\\
 \Delta m_{B_d}^{total} &\approx \Delta m_{B_d}^{0} + \Delta m_{B_d}^{1}  \approx
 \left[\frac{ |{\cal O}_{2_{bd}}|^2  }{m_{Z'_2}^2} + \frac{|{\cal O}_{3_{bd}}|^2  }{m_{Z'_3}^2}
 -\frac{ |F_{23}|^2 |F_{21}|^2}{3(4\pi)^2} F_{box}(m_S,M')   \right] m_{B_d} f^2_{B_d},\\
 \Delta m_{B_s}^{total} & \approx \Delta m_{B_s}^{0} + \Delta m_{B_s}^{1}   \approx
 \left[\frac{ |{\cal O}_{2_{bd}}|^2  }{m_{Z'_2}^2} + \frac{|{\cal O}_{3_{bd}}|^2  }{m_{Z'_3}^2}
 -\frac{ |F_{23}|^2 |F_{22}|^2}{3(4\pi)^2} F_{box}(m_S,M')   \right] m_{B_s} f^2_{B_s},
\label{eq:kk-total-bs}
\end{align}
where $\Delta m_K^0$ is negligibly small.
Thus, we do not need to consider the constraints of  the $M-\bar M$ mixings, since we can expect that the contributions  get canceled 
among Eqs.(\ref{eq:kk-total-k})-(\ref{eq:kk-total-bs}).\footnote{With $Q'$ only, 
the value of $\Delta C_9$ in Eq.~(\ref{eq:C9-fit}) can be achieved if
$M'\approx$ 300 GeV and $m_S\approx$ 100 GeV.
However, the lower mass bound for the exotic quark $Q'$ is of the order 1 TeV from the LHC~\cite{Sirunyan:2017kiw}.
 On the other hand,
  with $d'$ only, there is no solution to satisfy the constraint of $\Delta m_K$ within the perturbative $G$. 
  Discussions of the dark matter candidate $S$ can be found in ref.~\cite{Hutauruk:2019crc}. }

\section{Numerical analysis}
In our numerical analysis,
we explore the allowed  region of $-\Delta C_9$, by randomly selecting the  input parameters of $g'_{1,2,3}$ and $m_{Z'_{2,3}}$,
along with all the constraints discussed above. 
Then, each of the scan range is taken to be 
\begin{align}
& v'_{1,2,3} \in [10^3, 10^7]\ {\rm GeV}, \quad  g'_{2,3} \in [10^{-5}, \sqrt{4\pi}] .
\label{eq:setting2}
\end{align}
Fig.~\ref{fig:c9} shows  the possible regions in the planes of  $s_\theta$-$(-\Delta C_9)$  (left) and $g'_{2(3)}/m_{Z'_{2,3}}$-$(-\Delta C_9)$ (right),
where the horizontal  black (green) line corresponds to the observed value of 1.03 ($\pm$0.20), which is allowed by the experiment in Eq.~(\ref{eq:C9-fit}).
The figure at the left-handed side of Fig.~\ref{fig:c9} suggests that a larger $s_\theta$ is favored with the allowed lowest range being
about $0.2$.
%
The right-handed figure  in Fig.~\ref{fig:c9} indicates  that $g'_3/m_{Z'_3}$ does not depend on $-\Delta C_9$ so much, 
whereas $g'_2/m_{Z'_2}$ does, resulting in the allowed ranges of  $g'_3/m_{Z'_3}\lesssim0.10$ and $0.13\lesssim g'_2/m_{Z'_2}\lesssim0.16$. 

%
\begin{figure}[t]
\begin{center}
\includegraphics[width=60mm]{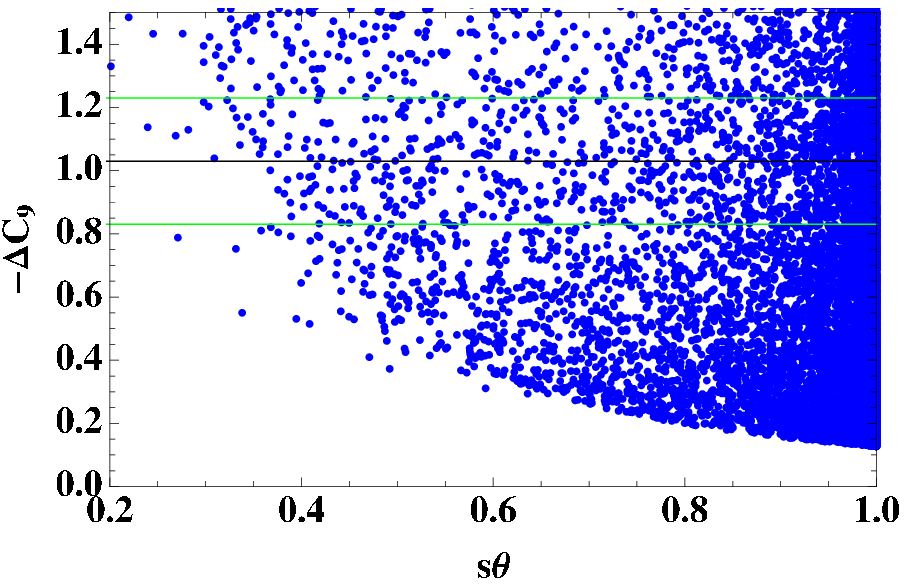} 
\includegraphics[width=60mm]{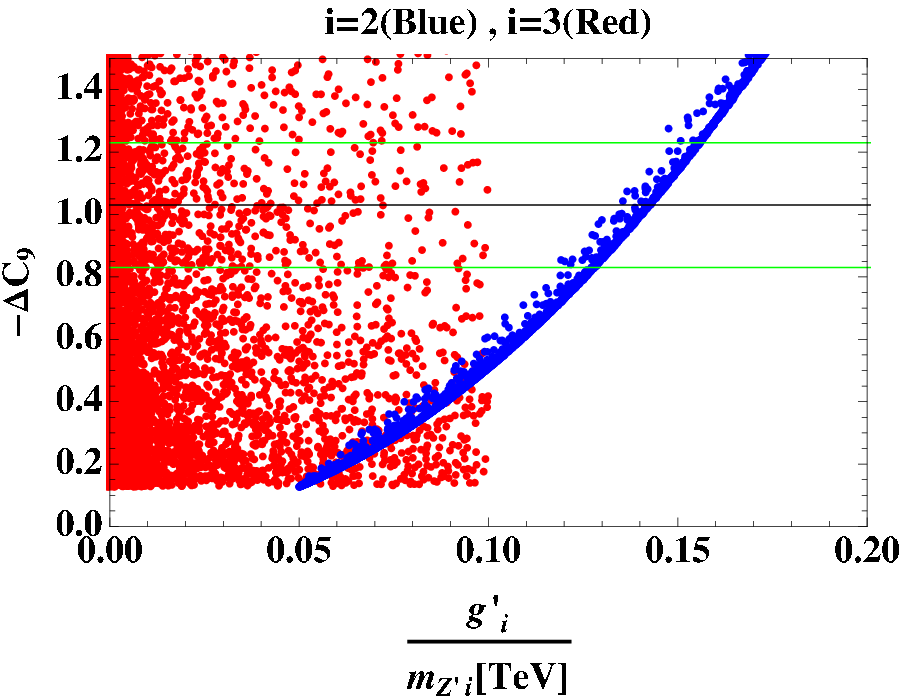} 
\caption{Allowed regions in the planes of  $s_\theta$-$(-\Delta C_9)$ (left)  and $g'_{2,3}/m_{Z'_{2,3}}$-$(-\Delta C_9)$ (right),
 where the horizontal  black (green) line corresponds to the observed value of 1.03 ($\pm$0.20), allowed by the experiment in Eq.~(\ref{eq:C9-fit}).}
  \label{fig:c9} 
\end{center}\end{figure}
%

\section{Conclusions and Discussions}
We have proposed a model with flavor dependent gauged symmetries of $U(1)_{B_1-L_1}\times U(1)_{B_2-L_2}\times U(1)_{B_3-L_3}$.
In this framework, we have formulated the renormalizable Yukawa Lagrangian, Higgs potential and kinetic term.
We have found that no additional Higgs boson is needed to avoid the dangerous GB, which is one of the main modification
of  the  model in Ref.~\cite{Mu:2018weh}.
Based on the successful two-zero texture, we are able to give several predictions in the lepton sector
as concretely shown in our numerical analysis.
We have also formulated the mass matrix in the additional neutral gauge bosons,
and successfully decomposed the electron/positron specific gauge boson and the others, imposing some assumptions.
Due to this decomposition, the strong constraint  from LEP experiment has been evaded.
This is also an improvement on the
model in Ref.~\cite{Mu:2018weh}.
Finally, we have done a global numerical analysis by
including all of the valid constraints, and illustrated the allowed region to satisfy the anomaly of 
$B\to K^*\bar\mu\mu$ via additional gauge bosons.

\section*{Acknowledgments}
This work was supported in part by National Center for Theoretical Sciences and
MoST (MoST-104-2112-M-007-003-MY3 and MoST-107-2119-M-007-013-MY3) (CQG), and
 the Ministry of Science, ICT and Future Planning, Gyeongsangbuk-do and Pohang City (H.O.). 

\end{document}